\documentclass{article}
\usepackage{spconf,amsmath,graphicx}
\usepackage{amssymb}
\usepackage{subfig}
\usepackage[hidelinks]{hyperref}
\usepackage{color}
\usepackage{siunitx}
\usepackage{booktabs}
\usepackage{etoolbox}
\usepackage{multirow}
\usepackage{amsmath}
\usepackage{changes}
\newrobustcmd{\B}{\bfseries}

\title{Improving Video Colorization by Test-Time Tuning}
\name{ Yaping Zhao$^{1,3,*}$
 , Haitian Zheng$^{2,*}$, Jiebo Luo$^{2}$, Edmund Y. Lam$^{1,3,\dag}$}

\address{
$^1$ The University of Hong Kong, Pokfulam, Hong Kong SAR \, \indent \,
$^2$ University of Rochester, USA\\
$^3$ACCESS –- AI Chip Center for Emerging Smart Systems, Hong Kong SAR}
	
\begin{document}
\newcommand{\Amat}{{\bf A}}
\newcommand{\Bmat}{{\bf B}}
\newcommand{\Cmat}{{\bf C}}
\newcommand{\Dmat}{{\bf D}}
\newcommand{\Emat}[0]{{{\bf E}}}
\newcommand{\Fmat}[0]{{{\bf F}}}
\newcommand{\Gmat}[0]{{{\bf G}}}
\newcommand{\Hmat}[0]{{{\bf H}}}
\newcommand{\Imat}{{\bf I}}
\newcommand{\Jmat}[0]{{{\bf J}}}
\newcommand{\Kmat}[0]{{{\bf K}}}
\newcommand{\Lmat}[0]{{{\bf L}}}
\newcommand{\Mmat}[0]{{{\bf M}}}
\newcommand{\Nmat}[0]{{{\bf N}}}
\newcommand{\Omat}[0]{{{\bf O}}}
\newcommand{\Pmat}[0]{{{\bf P}}}
\newcommand{\Qmat}[0]{{{\bf Q}}}
\newcommand{\Rmat}[0]{{{\bf R}}}
\newcommand{\Smat}[0]{{{\bf S}}}
\newcommand{\Tmat}[0]{{{\bf T}}}
\newcommand{\Umat}{{{\bf U}}}
\newcommand{\Vmat}[0]{{{\bf V}}}
\newcommand{\Wmat}[0]{{{\bf W}}}
\newcommand{\Xmat}{{\bf X}}
\newcommand{\Ymat}[0]{{{\bf Y}}}
\newcommand{\Zmat}{{\bf Z}}

\newcommand{\av}{\boldsymbol{a}}
\newcommand{\Av}{\boldsymbol{A}}
\newcommand{\Cv}{\boldsymbol{C}}
\newcommand{\bv}{\boldsymbol{b}}
\newcommand{\cv}{{\boldsymbol{c}}}
\newcommand{\dv}{\boldsymbol{d}}
\newcommand{\ev}[0]{{\boldsymbol{e}}}
\newcommand{\fv}{\boldsymbol{f}}
\newcommand{\Fv}[0]{{\boldsymbol{F}}}
\newcommand{\gv}[0]{{\boldsymbol{g}}}
\newcommand{\hv}[0]{{\boldsymbol{h}}}
\newcommand{\iv}[0]{{\boldsymbol{i}}}
\newcommand{\jv}[0]{{\boldsymbol{j}}}
\newcommand{\kv}[0]{{\boldsymbol{k}}}
\newcommand{\lv}[0]{{\boldsymbol{l}}}
\newcommand{\mv}[0]{{\boldsymbol{m}}}
\newcommand{\nv}{\boldsymbol{n}}
\newcommand{\ov}[0]{{\boldsymbol{o}}}
\newcommand{\pv}[0]{{\boldsymbol{p}}}
\newcommand{\qv}[0]{{\boldsymbol{q}}}
\newcommand{\rv}[0]{{\boldsymbol{r}}}
\newcommand{\sv}[0]{{\boldsymbol{s}}}
\newcommand{\tv}[0]{{\boldsymbol{t}}}
\newcommand{\uv}[0]{{\boldsymbol{u}}}
\newcommand{\vv}{\boldsymbol{v}}
\newcommand{\wv}{\boldsymbol{w}}
\newcommand{\Wv}{\boldsymbol{W}}
\newcommand{\xv}{\boldsymbol{x}}
\newcommand{\yv}{\boldsymbol{y}}
\newcommand{\Xv}{\boldsymbol{X}}
\newcommand{\Yv}{\boldsymbol{Y}}
\newcommand{\zv}{\boldsymbol{z}}

\newcommand{\Gammamat}[0]{{\boldsymbol{\Gamma}}}
\newcommand{\Deltamat}[0]{{\boldsymbol{\Delta}}}
\newcommand{\Thetamat}{\boldsymbol{\Theta}}
\newcommand{\Lambdamat}{{\boldsymbol{\Lambda}}}
\newcommand{\Ximat}[0]{{\boldsymbol{\Xi}}}
\newcommand{\Pimat}[0]{{\boldsymbol{\Pi}} }
\newcommand{\Sigmamat}{\boldsymbol{\Sigma}}
\newcommand{\Upsilonmat}[0]{{\boldsymbol{\Upsilon}} }
\newcommand{\Phimat}{\boldsymbol{\Phi}}
\newcommand{\Psimat}{\boldsymbol{\Psi}}
\newcommand{\Omegamat}{{\boldsymbol{\Omega}}}

\newcommand{\Lambdav}{\bm{\Lambda}}
\newcommand{\alphav}{\boldsymbol{\alpha}}
\newcommand{\betav}[0]{{\boldsymbol{\beta}} }
\newcommand{\gammav}{{\boldsymbol{\gamma}}}
\newcommand{\deltav}[0]{{\boldsymbol{\delta}} }
\newcommand{\epsilonv}{\boldsymbol{\epsilon}}
\newcommand{\zetav}[0]{{\boldsymbol{\zeta}} }
\newcommand{\etav}[0]{{\boldsymbol{\eta}} }
\newcommand{\thetav}{\boldsymbol{\theta}}
\newcommand{\iotav}[0]{{\boldsymbol{\iota}} }
\newcommand{\kappav}{{\boldsymbol{\kappa}}}
\newcommand{\lambdav}[0]{{\boldsymbol{\lambda}} }
\newcommand{\muv}{\boldsymbol{\mu}}
\newcommand{\nuv}{{\boldsymbol{\nu}}}
\newcommand{\xiv}{{\boldsymbol{\xi}}}
\newcommand{\omicronv}[0]{{\boldsymbol{\omicron}} }
\newcommand{\piv}{\boldsymbol{\pi}}
\newcommand{\rhov}[0]{{\boldsymbol{\rho}} }
\newcommand{\sigmav}[0]{{\boldsymbol{\sigma}} }
\newcommand{\tauv}[0]{{\boldsymbol{\tau}} }
\newcommand{\upsilonv}[0]{{\boldsymbol{\upsilon}} }
\newcommand{\phiv}{\boldsymbol{\phi}}
\newcommand{\chiv}[0]{{\boldsymbol{\chi}} }
\newcommand{\psiv}{\boldsymbol{\psi}}
\newcommand{\omegav}[0]{{\boldsymbol{\omega}} }

\newcommand{\xin}[1]{{\textcolor{red}{#1}}}

\newcommand{\ts}{^{\top}}
\newcommand{\TV}{{\rm TV}}
\newtheorem{definition}{Definition}
\newtheorem{lemma}{Lemma}
\newtheorem{corollary}{Corollary}
\newtheorem{theorem}{Theorem}%[section]

\twocolumn[{%
\renewcommand\twocolumn[1][]{#1}%
\maketitle
\begin{center}
\vspace{-20pt}
    \includegraphics[width=1.\linewidth]{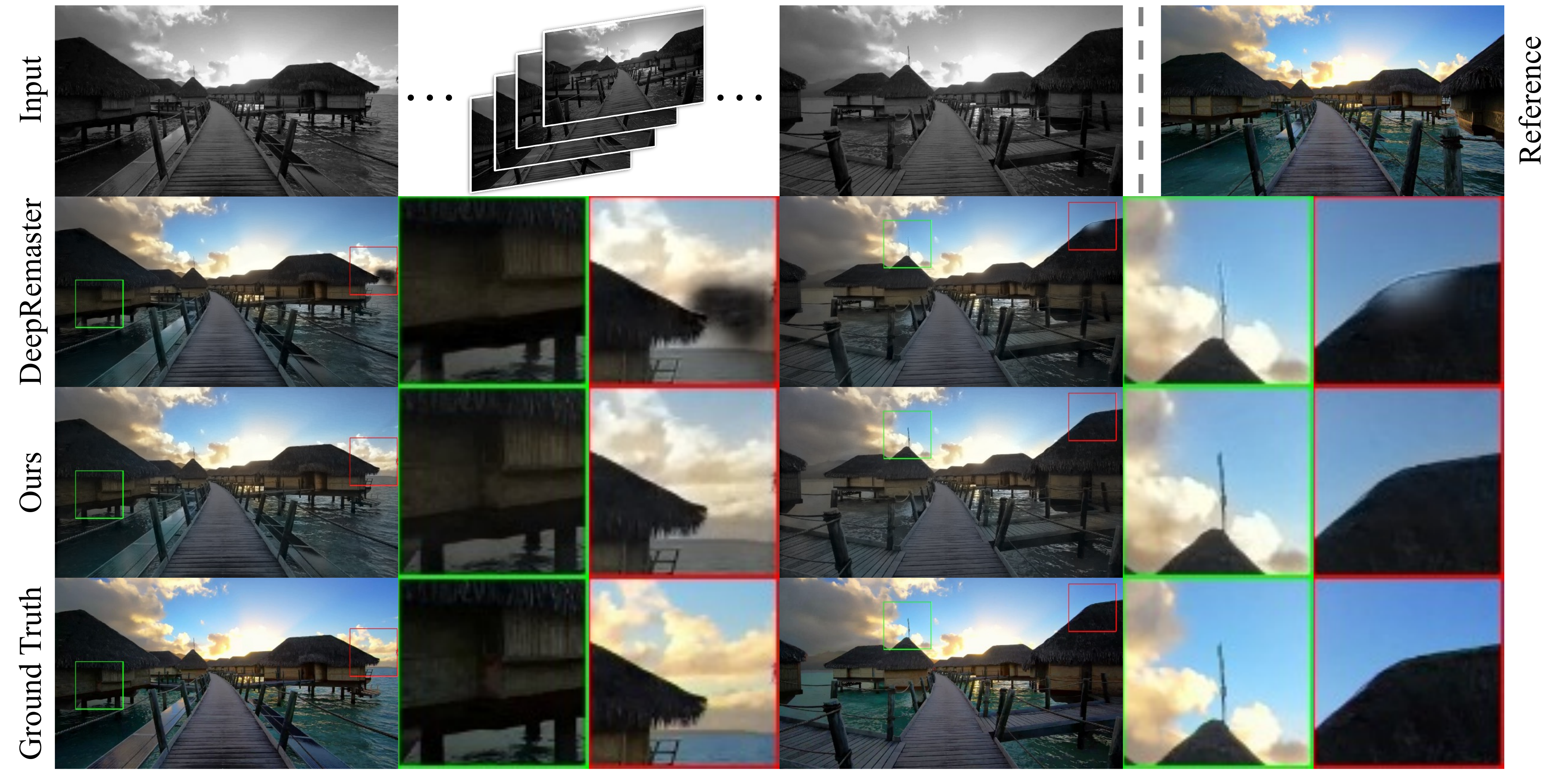}
    \vspace{-15pt}
	\captionof{figure}{
Given a monochrome video sequence as input and the colorized first frame as reference, we propose a simple yet effective method using test-time tuning on pretrained model DeepRemaster~\cite{IizukaSIGGRAPHASIA2019} to perform video colorization with superior performance. Green and red boxes are zooming into details.
% Our proposed method, named $T^3$, aims to perform video colorization given a monochrome video sequence as input and the colorized first frame as reference.
% test-time tuning on a pre-trained model, DeepRemaster~\cite{IizukaSIGGRAPHASIA2019}, using a colorized first frame as reference. We demonstrate that our approach achieves superior performance with simple and effective techniques. We provide zoomed-in details of colorized frames with green and red boxes.
}
	\label{fig:teaser}
	\vspace{10pt}
\end{center}
}
]

\newcommand\blfootnote[1]{%
  \begingroup
  \renewcommand\thefootnote{}\footnote{#1}%
  \addtocounter{footnote}{-1}%
  \endgroup
}

\blfootnote{$^{*}$Equal contribution.  $^{\dag}$Corresponding author.}
\blfootnote{This work is supported in part by the Research Grants Council (GRF 17201620), by the Research Postgraduate Student Innovation Award (The University of Hong Kong), and by ACCESS –- AI Chip Center for Emerging Smart Systems, Hong Kong SAR.}

% \maketitle

% \begin{figure*}
%     \includegraphics[width =\linewidth]{img/teaser1.png}
%     \caption{Illustration of the human-centric RefSR that transfers the high-definition human body details onto low resolution video.}
%     \label{fig:teaser}
%     \vspace{-10pt}
% \end{figure*}
\vspace{-5mm}
\begin{abstract}
With the advancements in deep learning, video colorization by propagating color information from a colorized reference frame to a monochrome video sequence has been well explored. However, the existing approaches often suffer from overfitting the training dataset and sequentially lead to suboptimal performance on colorizing testing samples. To address this issue, we propose an effective method, which aims to enhance video colorization through test-time tuning. By exploiting the reference to construct additional training samples during testing, our approach achieves a performance boost of $1\sim3$ \si{\decibel} in PSNR on average compared to the baseline.
Code is available at: \href{https://github.com/IndigoPurple/T3}{\textcolor{blue}{https://github.com/IndigoPurple/T3}}.
% \hz{please briefly describe the general idea and the performance.}
% Experiments demonstrate the effectiveness and efficiency of our approach.

\end{abstract}


\begin{figure*}
    \centering
    \includegraphics[width =\linewidth]{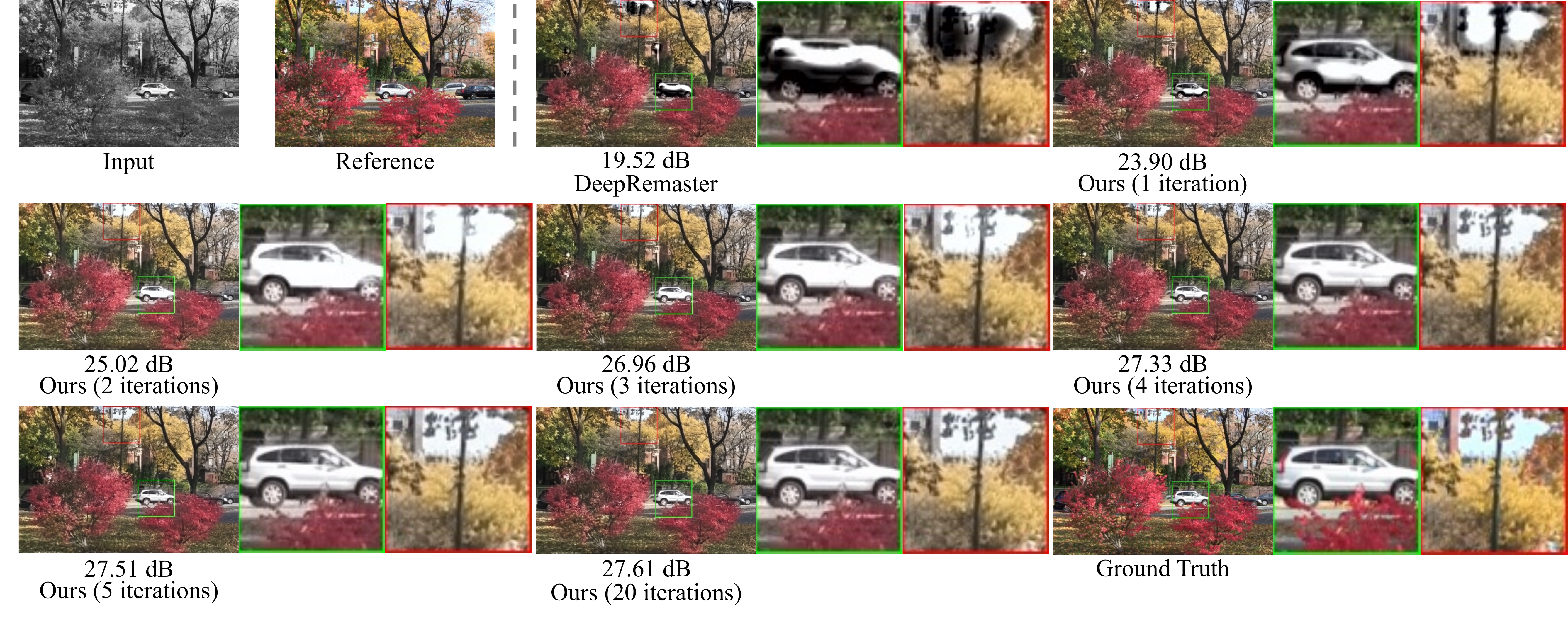}
    \caption{Video colorization comparisons on the video sequence \texttt{foliage} of the Vid4 dataset~\cite{liu2013bayesian}. While DeepRemaster~\cite{IizukaSIGGRAPHASIA2019} severely results in artifacts and distortion, our method boosts the performance with merely a few iterations for test-time tuning. Quantitative evaluations are provided in terms of the PSNR metric. Green and red boxes are zooming into details.}
    \label{fig:iter}
\end{figure*}

\section{introduction}
\label{sec:intro}
Nowadays, it has become common practice to leverage an ideally colorized video frame as a reference to colorize an entire monochrome video sequence. Suah a video colorization task has a wide range of applications, such as reviving vintage films. State-of-the-art methods for video colorization rely on deep learning and benefit from large-scale datasets. However, these existing approaches inevitably suffer from overfitting the training dataset, which in turn leads to suboptimal performance on colorizing testing samples in a sequential manner.

To address this issue, our paper introduces a simple yet effective method that enhances video colorization through test-time tuning. Our key insight is that the reference can be considered as ground truth and can help construct an additional training sample during testing. Specifically, we demonstrate that using the "ref-mono" pair, which comprises the reference with color information and its corresponding monochrome video frame, is adequate to guide the pre-trained model tuning towards higher performance.

To demonstrate this, we utilize the pre-trained model from DeepRemaster~\cite{IizukaSIGGRAPHASIA2019} and fine-tune the network parameters with objective functions that only consider the ref-mono pair. By doing so, we expose testing samples that are previously unseen in the training dataset to the neural network for adaptive tuning. Our experiments on various datasets showcase the effectiveness and efficiency of our proposed method.

Our main contributions are as follows:
\begin{itemize}
    % \item A \textbf{simple yet effective} method to improve video colorization by test-time tuning, which can boost the performance by \SI{3}{\decibel} compared to the baseline.
    \item A \textbf{simple yet effective} method that improves video colorization through test-time tuning, resulting in a performance boost of $1\sim3$ \si{\decibel} to the baseline.
    % \item A \textbf{low-cost} finetuning paradigm that utilizes the reference to finetune pre-trained models, without incurring more network parameters or requiring additional labels.
    \item A \textbf{low-cost} fine-tuning paradigm that utilizes the reference to finetune pre-trained models, without requiring additional network parameters or annotated labels.
    % \item A \textbf{high-efficiency} iterative optimization that updates the network parameters to achieve high performance with a few iterations and thus a short processing time.
    \item A \textbf{highly efficient} iterative optimization that updates the network parameters to achieve high performance within a few iterations, making it feasible for real-time applications.
\end{itemize}

\section{related work}
\label{sec:related}
In video colorization, one or more video frames with color information are typically provided as reference to provide cues and ensure accuracy. While image colorization methods~\cite{he2018deep,xiao2020example,zhao2022cross} can be directly applied to video colorization, Zhang \textit{et al.}~\cite{zhang2019deep} introduced a temporal consistency loss~\cite{chen2017coherent} to improve video colorization with a recurrent framework. Later, video colorization is further developed by incorporating GAN encoders~\cite{wu2021towards} and visual tracking~\cite{vondrick2018tracking}. DeepRemaster~\cite{IizukaSIGGRAPHASIA2019} established source-reference correspondence by finding similarities within the reference image and target image, which has been frequently used in recent years~\cite{lee2020reference,li2020zoom,zhao2021cross,siyao2021deep,zhao2021efenet,shi2022reference,zhao2022manet}.
 
% To colorize a monochrome video sequence, usually one or more video frames with color information are given as reference to provide the cues and guarantee the correctness. While image colorization methods~\cite{he2018deep,xiao2020example,zhao2022cross} could be directly applied to video colorization, Zhang \textit{et al.}~\cite{zhang2019deep}  introduced a temporal consistency loss~\cite{chen2017coherent} to improve video colorization with a recurrent framework. Later, video colorization is further developed by incorporating generative adversarial network (GAN) encoders~\cite{wu2021towards} and visual tracking~\cite{vondrick2018tracking}. DeepRemaster~\cite{IizukaSIGGRAPHASIA2019} is proposed to introduce source-reference attention by finding the non-local similarities within the extracted reference image and target image features, which has been frequently used in recent years~\cite{lee2020reference,siyao2021deep,zhao2021efenet,shi2022reference,zhao2022manet}.

\begin{table*}[]
	\centering
	\resizebox{\textwidth}{!}{
\begin{tabular}{c|cccccccc|c}
    \toprule
        Method & tractor & motorbike & sunflower & park\_joy & snowboard & touchdown & rafting & hypersmooth & Average \\
        \midrule
        Zhang \textit{et al.}~\cite{zhang2019deep} & 18.22, 0.8518 & 23.05, 0.9043 & 15.50, 0.7376 & 21.36, 0.8707 & 19.30, 0.9348 & 25.44, 0.9144 & 22.54, 0.9105 & 23.73, 0.9141 & 21.14, 0.8798\\
        DeepRemaster~\cite{IizukaSIGGRAPHASIA2019} & 24.09, 0.8791 & 24.51, 0.9024 & 24.57, 0.8268 & 25.22, 0.8627 & 24.07, 0.9349 & 27.25, 0.9270 & 23.41, 0.9133 & 26.05, 0.9074 & 24.90, 0.8942\\
        Ours & \textbf{25.11, 0.8867} & \textbf{25.36, 0.9077} & \textbf{24.82, 0.8290} & \textbf{26.02, 0.8878} & \textbf{24.79, 0.9440} & \textbf{30.19, 0.9329} & \textbf{24.47, 0.9151} & \textbf{26.52, 0.9190} & \textbf{25.91, 0.9028}\\
        \bottomrule
\end{tabular}
}
	\caption{Video colorization comparisons on the Set8 dataset~\cite{tassano2019dvdnet} in terms of PSNR and SSIM.
	}
\label{tab:set8}
\end{table*}

\begin{figure*}
    \centering
    \includegraphics[width =\linewidth]{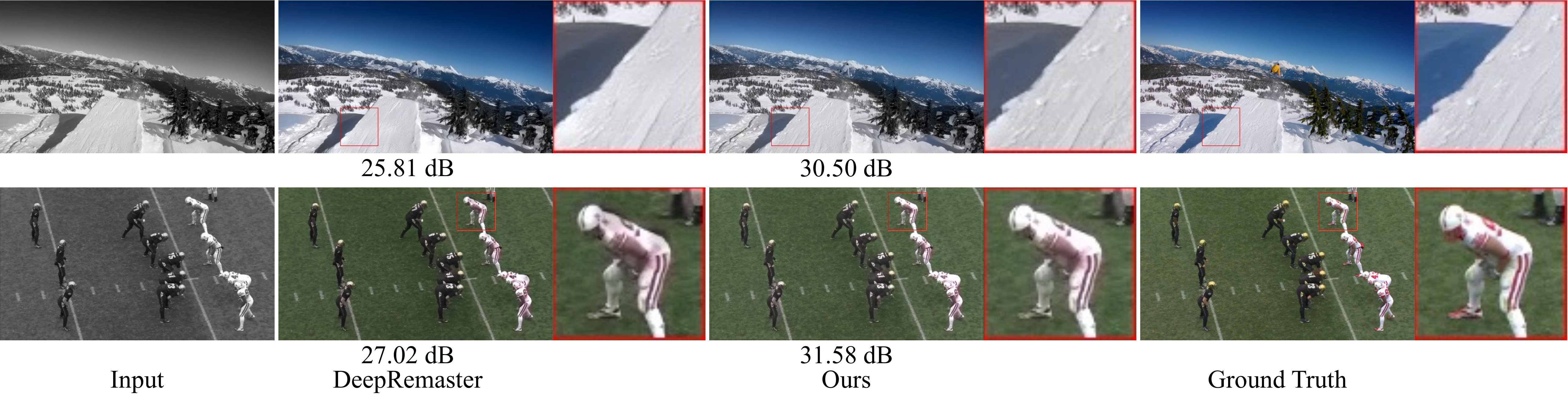}
    \caption{Video colorization comparisons on the video sequence \texttt{snowboard} (the upper row) and \texttt{touchdown} (the bottom row) of the Set8 dataset~\cite{tassano2019dvdnet}. While DeepRemaster~\cite{tassano2019dvdnet} suffers in detail loss and artifacts, our method reconstructs explicit content. Quantitative evaluations are provided in terms of the PSNR metric. Red boxes are zooming into details.}
    \label{fig:set8}
\end{figure*}

\section{method}
\label{sec:method}

\subsection{Overall Framework}
Given a monochrome video sequence as input and the colorized first frame as reference, we denote the input as $\Xmat = \{ \xv^{(t)} \}_{t=1}^T$ and the reference as $\zv$, where $t$ indicates the time step of the video frame, and $T$ is the total number of video frames. Moreover, we assume $\xv \in {\mathbb{R}}^{H \times W \times 1}$, $\zv \in {\mathbb{R}}^{H \times W \times 3}$, where $H$ and $W$ denote the height and width, 1 and 3 are the channel numbers for grayscale (input) or RGB (reference).

To perform video colorization, deep networks are applied by implicitly learning a function $\hat{\Ymat} = f_\theta(\Xmat; \zv)$ that maps a monochrome video sequence $\Xmat$ to a colorized one $\hat{\Ymat} = \{ {\hat{\yv}}^{(t)}\}_{t=1}^T$, where $\theta$ is the network parameters we aim to optimize, and ${\hat{\yv}}^{(t)} \in {\mathbb{R}}^{H \times W \times 3}$ is the synthesized video frame at time step $t$. 
However, state-of-the-art neural networks for video colorization are all invariably trained on datasets. While they learn priors from the dataset to achieve excellent performance, they inevitably suffer in overfitting the training samples,
which usually have a gap with the testing samples. Sequentially and unfortunately, such a gap leads to suboptimal performance. 

To address this issue, we propose a test-time tuning approach that leverages an ideally colorized reference frame $\zv$. The pre-trained network is fine-tuned using an combined objective function, which improve color transfer from the reference to the monochrome. 

\subsection{Objective Function}
Our aim is to alleviate the overfitting problem by test-time tuning the pre-trained network parameters. To achieve this, we make full use of the reference, which can be considered ideally colorized, \textit{i.e.}, ground truth. Our key insight is that the reference can provide a label to construct an additional training sample during testing time.
Specifically, we adopt the neural network from DeepRemaster~\cite{IizukaSIGGRAPHASIA2019} and employ an objective function $\mathcal{L}$ to perform finetuning on the pre-trained model. Considering the ref-mono pair $\hat{\yv}^{(1)}$ and $\zv$, though we can directly apply a loss function like $\mathcal{L}_{rgb} = || \hat{\yv}^{(1)} - \zv ||_2^2$, to finetune the model parameters, we use a well-designed set of combined objective functions using LAB color space: 
\begin{align}
    \mathcal{L} &= \mathcal{L}_l + \mathcal{L}_{ab},
\end{align}
where the overall objective $\mathcal{L}$ for finetuning regularization is consist of two loss functions $\mathcal{L}_l$ and $\mathcal{L}_{ab}$. The former prevents the neural network to generate artifacts and noise. The latter is responsible for better transferring the colors from the reference to the monochrome. Specifically, $\mathcal{L}_l$ and $\mathcal{L}_{ab}$ are calculated by:
\begin{align}
    \mathcal{L}_l &= || \hat{\yv}^{(1)}_l - \zv_l ||_2^2,\label{eq:l_l}\\
    \mathcal{L}_{ab} &= || \hat{\yv}^{(1)}_{ab} - \zv_{ab} ||_2^2,
    \label{eq:l_ab}
\end{align}
where $\hat{\yv}^{(1)}_l$ and $\hat{\yv}^{(1)}_{ab}$ are the luminance and chrominance component, respectively, of the first colorized video frame $\hat{\yv}^{(1)}$ in the LAB color space. Similarly, the $\zv_l$ and $\zv_{ab}$ are obtained by splitting the reference $\zv$ to the luminance and chrominance.

By simply using our proposed objective functions, our approach effectively improve the performance of the existing method DeepRemaster~\cite{IizukaSIGGRAPHASIA2019} in a test-time tuning manner with only a few iterations. As shown in Figure~\ref{fig:iter}, while DeepRemaster~\cite{IizukaSIGGRAPHASIA2019} often results in artifacts and distortion,  our method improves the PSNR performance by more than \SI{4}{\decibel} after only one iteration of test-time tuning and by \SI{8}{\decibel} after only 20 iterations. 

\begin{table*}[]
    \centering
    \begin{tabular}{c|cccc|c}
    \toprule
        Method & Calendar & City & Foliage & Walk & Average \\
        \midrule
        Zhang \textit{et al.}~\cite{zhang2019deep} & 18.98, 0.9126 & 32.18, 0.9676 & 19.13, 0.9181 & 26.17, 0.9507 &24.12, 0.9372\\
        DeepRemaster~\cite{IizukaSIGGRAPHASIA2019} & 23.39, 0.9127 & 28.49, 0.9662 & 19.52, 0.9220 & 27.16, 0.9504 & 24.64, 0.9378 \\ 
        Ours & \textbf{23.81, 0.9139} & \textbf{33.06, 0.9725} & \textbf{26.47, 0.9441} & \textbf{28.10, 0.9544} & \textbf{27.86, 0.9462}\\
        \bottomrule
    \end{tabular}
    \caption{Video colorization comparisons on the Vid4 dataset~\cite{liu2013bayesian} in terms of PSNR and SSIM.}
    \label{tab:vid4}
\end{table*}

\begin{figure*}
    \centering
    \includegraphics[width =\linewidth]{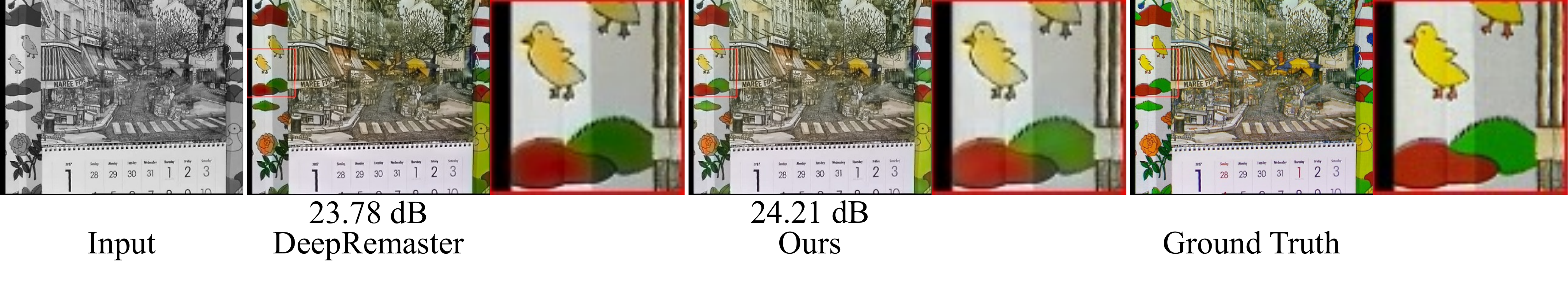}
    \caption{Video colorization comparisons on the video sequence \texttt{calendar} of the Vid4 dataset~\cite{liu2013bayesian}. Compared to DeepRemaster~\cite{tassano2019dvdnet}, the colors of our results are more consistent with the reference. Quantitative evaluations are provided in terms of the PSNR metric. Red boxes are zooming into details.}
    \label{fig:vid4}
\end{figure*}

\begin{table}[]
    \centering
    \begin{tabular}{c|c|c|c}
    \toprule
        Resolution & $320\times576$ & $320\times400$ & $256\times256$\\
        \midrule
        Time (5 iters) & 1.50 & 1.13 & 0.53\\
        Time (20 iters) & 5.47 & 4.10 & 1.94\\
        \bottomrule
    \end{tabular}
    \caption{Average tuning time per video sequence in minutes using identical environments, regarding different image resolutions and optimization iterations. The image resolution is denoted as $H \times W$, where $H$ and $W$ are height and width, respectively.}
    \label{tab:time}
\end{table}

% \begin{table}[]
%     \centering
%     \begin{tabular}{c|c|c|c}
%     \toprule
%         Resolution & $576\times320$ & $400\times320$ & $256\times256$\\
%         \midrule
%         Time & 16.41 & 12.31 & 5.83\\
%         \bottomrule
%     \end{tabular}
%     \caption{Average tuning time per video sequence in minutes on PU-Net~\cite{yu2018pu} dataset using identical environments.}
%     \label{tab:time}
% \end{table}

\section{experiment}
\label{sec:exp}

% \subsection{Experimental Setting}
\noindent{\textbf{Experiment Setting.}}
We evaluated our method on two commonly used video datasets, Vid4~\cite{liu2013bayesian} and Set8~\cite{tassano2019dvdnet}, using both quantitative and qualitative metrics. For quantitative evaluation, we used two commonly used metrics, PSNR and SSIM~\cite{wang2004image}. We compared our method to two video colorization methods, Zhang~\textit{et al.}\cite{zhang2019deep} and DeepRemaster\cite{IizukaSIGGRAPHASIA2019}. In our experiments, the pre-trained model from DeepRemaster~\cite{IizukaSIGGRAPHASIA2019} is finetuned for $20$ iterations using the Adam~\cite{kingma2014adam} optimizer, with a learning rate of $1 \times 10^{-4}$. We experimented with other iteration steps and found that increasing the number of iterations beyond 20 did not significantly improve PSNR, with a saturation effect observed. For example, increasing the number of iterations to 50 only increased PSNR by approximately 0.07 compared to 20 iterations.
\newline
% Two commonly used video datasets, Vid4~\cite{liu2013bayesian} and Set8~\cite{tassano2019dvdnet}, are adopted for both quantitative and qualitative evaluations. Two commonly used metrics, PSNR and SSIM~\cite{wang2004image}, are employed for quantitative evaluations. Two video colorization methods, Zhang~\textit{et al.}~\cite{zhang2019deep} and DeepRemaster~\cite{IizukaSIGGRAPHASIA2019}, are used for comparison. In our experiments, the pre-trained model from DeepRemaster~\cite{IizukaSIGGRAPHASIA2019} is finetuned for $20$ iterations using the Adam~\cite{kingma2014adam} optimizer. The learning rates are set to $1 \times 10^{-4}$. We experimented with other iteration steps and found that PSNR saturates after 20 iterations, \textit{e.g.}, 50 iterations increase PSNR by approximately $0.07$ compared to 20 iterations.
% \subsection{Quantitative and Qualitative Results}
\newline
\noindent{\textbf{Quantitative and Qualitative Results.}}
As Table~\ref{tab:set8} and~\ref{tab:vid4} show, our model outperforms previous methods Zhang~\textit{et al.}~\cite{zhang2019deep} and DeepRemaster~\cite{IizukaSIGGRAPHASIA2019} on all the video sequences of different datasets. 
To further illustrate the effectiveness of our method, we compare the qualitative performance with the competitive baseline DeepRemaster~\cite{IizukaSIGGRAPHASIA2019}. 
As Figure~\ref{fig:set8} shows, while DeepRemaster~\cite{tassano2019dvdnet} suffers from artifacts and loss of details, our method reconstructs explicit contents.
In Figure~\ref{fig:vid4}, compared to DeepRemaster~\cite{tassano2019dvdnet}, the colors of our results are more consistent with the reference.
\newline
% \subsection{Inference Time}
\noindent{\textbf{Tuning Time.}}
As Table~\ref{tab:time} shows, our method requires a short time for tuning the model. Moreover, we observe that the PSNR could achieve a relatively higher level after merely 5 iterations, which is also intuitively verified in Figure~\ref{fig:iter}. Therefore, to reduce tuning time, the iteration number could be set to 5. 
\newline
% \subsection{Ablation Study}
\noindent{\textbf{Ablation Study.}}
We investigate the role of the loss functions formulated in Equation~\ref{eq:l_l} and \ref{eq:l_ab} by two variants of our pipeline denoted as Ours \textit{w/o} $\mathcal{L}_{l}$ and Ours \textit{w/o} $\mathcal{L}_{ab}$, which respectively turn off one of the components while remaining another.
In addition, we also explore the experimental results simply using $\mathcal{L}_{rgb} = || \hat{\yv}^{(1)} - \zv ||_2^2$, denoted as Baseline \textit{w/} $\mathcal{L}_{rgb}$. According to Table~\ref{tab:as}, disabling either component in LAB color space or using the $\mathcal{L}_{rgb}$ in RGB color space leads to a performance drop, suggesting the soundness of our proposed method. 

\begin{table}[]
    \centering
    \begin{tabular}{c|c|c}
    \toprule
        Dataset & Vid4~\cite{liu2013bayesian} & Set8~\cite{tassano2019dvdnet}\\
        \midrule
        Baseline \textit{w/} $\mathcal{L}_{rgb}$ & 27.65, 0.9421 & 25.63, 0.8974\\
        Ours \textit{w/o} $\mathcal{L}_{l}$ & 25.64, 0.9397 & 25.01, 0.8962\\
        Ours \textit{w/o} $\mathcal{L}_{ab}$ & 27.35, 0.9408 & 25.40, 0.8972\\
        Ours & \textbf{27.86, 0.9462} & \textbf{25.91, 0.9028}\\
        \bottomrule
    \end{tabular}
    \caption{ Ablation study regarding different objective functions for test-time tuning. The performance is evaluated in terms of average PSNR and SSIM on each dataset.}
    \label{tab:as}
\end{table}

\vspace{-2mm}
\section{Conclusion}
\vspace{-2mm}

In this paper, 
% we present a straightforward yet efficient approach for enhancing video colorization through test-time tuning.
% Our key insight relies on the fact that the reference can be viewed as the ground truth and can be used to generate an additional training sample during testing stage.
% Specifically, we show that merely using the reference with color information and its corresponding monochrome video frame, is sufficient to guide the pre-trained model in achieving higher performance after test-time tuning. 
% To show it, we employ the pre-trained model from recent work and finetune the network parameters, with a linear combination of objective functions using LAB color space. In this way, the testing samples previously unseen in the training dataset is disclosed to the neural network for adaptive tuning. Experiments on different datasets demonstrate the effectiveness and efficiency of our proposed method. 
we propose an efficient approach for enhancing video colorization through test-time tuning, leveraging the reference as a ground truth to generate an additional training sample during testing. By using the reference with color information and its corresponding monochrome frame, we show that the pre-trained model achieves higher performance after test-time tuning. We employ a pre-trained model from recent work and finetune the network parameters using a linear combination of objective functions with LAB color space. This enables adaptive tuning to previously unseen testing samples. Experimental results on various datasets validate the effectiveness and efficiency of our approach. For future work, another direction worthy of exploration is adding some noise to the network parameters before finetuning to alleviate the overfitting problem and further improve testing performance~\cite{wu2022noisytune,zhao2022improving}. 

\bibliographystyle{IEEEbib}
\bibliography{refs}

\end{document}